# High frequency Scanning Gate Microscopy and local memory effect of carbon nanotube transistors

Cristian Staii[1,*], Rui Shao[2], Dawn A. Bonnell[2] and Alan T. Johnson Jr.[1]

[1]*Department of Physics and Astronomy and Laboratory for Research on the Structure of Matter, University of Pennsylvania, Philadelphia, Pennsylvania 19104*

[2]*Department of Material Science and Engineering, University of Pennsylvania, Philadelphia, Pennsylvania 19104*

(Dated April 21, 2005)

We use impedance spectroscopy to measure the high frequency properties of single-walled carbon nanotube field effect transistors (swCN-FETs). Furthermore, we extend Scanning Gate Microscopy (SGM) to frequencies up to 15MHz, and use it to image changes in the impedance of swCN-FET circuits induced by the SGM-tip gate. In contrast to earlier reports, the results of both experiments are consistent with a simple RC parallel circuit model of the swCN-FET, with a time constant of 0.3 µs. We also use the SGM tip to show the local nature of the memory effect normally observed in swCN-FETs, implying that nanotube-based memory cells can be miniaturized to dimensions of the order of tens of nm.

Successful implementation of single-walled carbon nanotubes field effect transistors (swCN-FETs) as nanoelectronic circuit elements requires reliable techniques for the characterization of their local electronic properties and structure. At low frequencies, this goal was achieved by using recently developed scanning probe techniques.[1-5] In particular Scanning Gate Microscopy (SGM), and Scanning Impedance Microscopy (SIM) are powerful approaches for measuring the local electrical properties of swCN-FET devices. SGM was used to image Schottky barriers that develop within swCN-FET circuits,[2] and to quantify the local Fermi energy at gate-susceptible defects along the nanotube length.[3] SIM was used to measure the voltage distribution along the swCN-FET.[3] The extension of these techniques to high frequency is important because, although the low frequency performance of swCN-FET has been greatly improved,[6-9] little is known about their high frequency behavior.[10,11] In particular, although is has been shown[10] that swCN-FET can operate up to frequencies of the order of hundreds of megahertz, no information is available concerning the role of individual defects at frequencies higher than tens of kilohertz.

Here we present impedance characteristics of swCN-FET devices for ac-frequencies up to 15 MHz. We also combine SGM with Impedance Spectroscopy (IS) in order to extend the frequency range of SGM to 15 MHz. We use this High Frequency Scanning

Gate Microscopy (HF-SGM) technique to image changes in the impedance of a swCN-FET circuit induced by a voltage-biased AFM tip that acts as a nanoscale local gate. We show that the data are consistent with a parallel RC model for the swCN-FET circuit. We also use the tip gate to show that charge injection from the swCN-FET into the substrate, which is responsible for the memory effect, can be induced at specific locations along the nanotube length, implying that the active memory region in these devices may be only tens of nanometers in extent.

SwCN-FET circuits are fabricated from swCNs, grown directly on the substrate by catalytic chemical vapor deposition .[6] The substrate for all devices is a 200nm $SiO_2$ layer on top of a p-type degenerately doped Si wafer. Au electrodes are patterned using a shadow mask evaporation technique.[12] Circuits consisting of individual p-type semiconducting nanotubes were selected by using only devices that showed a strong decrease in the source-drain current $I_{SD}$ for positive back gate voltage $V_G$ at room temperature (ON/OFF ratio exceeding 1000). Such swCN-FETs all exhibited a large transconductance of order 1μS (inset Figure 1). These values indicate that Au makes excellent contacts to swCN-FETs.[13]

We begin by showing typical impedance spectra for a swCN-FET (Figure 1), taken using a HP429A impedance analyzer.[14] Figure 1 shows the variation of the impedance modulus |Z| with the ac frequency f in the 40Hz-15MHz interval, for two different backgate voltages: $V_G$= +10 V (green squares), and $V_G$=0 V (red circles). The blue triangles are data points taken with $V_G$=0 V, but with an SGM-tip gate placed at height h=15 nm above a particular location on the nanotube, and $V_{tip}$=+8 V (to be discussed below).

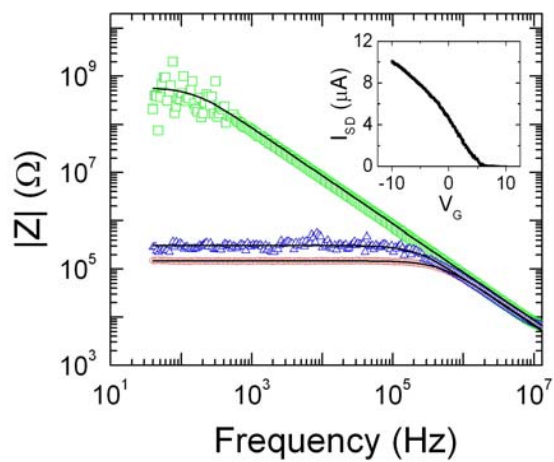

**Figure 1.** (a) Impedance spectra for backgate voltage $V_G$=+10V (green squares) and $V_G$=0V (red circles). Blue triangles are data taken with $V_G$ = 0 V and an SGM-tip positioned at h=15nm above the top defect seen in Figure 2b with $V_{tip}$= +8 V. Black curves are fits to the data using a parallel RC model (see text). Inset: Transport current $I_{SD}(V_G)$ for 1 V source-drain bias voltage.

The data in Figure 1 are in excellent agreement with a parallel RC circuit model for the swCN-FET, which predicts that $|Z(f)|=R/(1+(2\pi fRC)^2)^{1/2}$. Fits to the data (solid lines in Figure 1) give the following circuit parameters: $R_1$= 613MΩ, $C_1$ = 2.3 pF (top curve, $V_G$=+10V) and $R_2$=127KΩ, $C_2$ = 2.3 pF (bottom curve, $V_G$=0V). Identical values for R and C are obtained from fits to the measured phase of the impedance θ(f) (data not shown). DC measurements of $R(V_G)$ (Inset Figure 1) agree well with $R_1$ and $R_2$ found from the fits to the impedance spectra. The total capacitance of the circuit found from the fits to the data is not influenced by the backgate voltage. This is explained by the fact that the capacitance of the swCN-FET alone (calculated in the coaxial-cable approximation[6]) is of order 40aF; the circuit capacitance is therefore dominated by parasitics due to the metallic electrodes and wires. To test this conclusion, we measured Z(f) for circuits with identical contact pads and electrical connections as the swCN-FETs but *without* a nanotube between the electrodes. The capacitance of these test circuits (2-3 pF) was the same as that found for the swCN-FETs. This straightforward interpretation of Z(f) data is in contrast with previous reports of anomalous behavior, including a negative capacitance, in samples consisting of swCN bundles.[11]

To investigate the role of single defects in electronic transport through swCN-FET at high frequencies, we combine IS with SGM, thereby extending SGM to frequencies as large as 15MHz. Resonances in the cable connections limited the useful frequency range explored in these first measurements, so it should prove possible to extend this method to 1 GHz or higher in the future.

We begin with a discussion of low frequency (10kHz) SGM data (Figure 2b), taken on a Digital Instruments Dimension 3000 NS IIIA using $W_2C$ - coated tips (NSC12/W2C, Mikromasch). In SGM imaging mode,[1-3] a conducting tip with applied voltage $V_{tip}$ is scanned at a fixed height over an electrically biased sample, and the current $I_{SD}$ is recorded as a function of tip position. In contrast to a static backgate that couples capacitively to the entire sample, the tip is a spatially localized gate whose position can be varied. The image formed from $I_{SD}$ as a function of tip position reveals precise locations where the swCN-FET has a strong response to the tip gate.

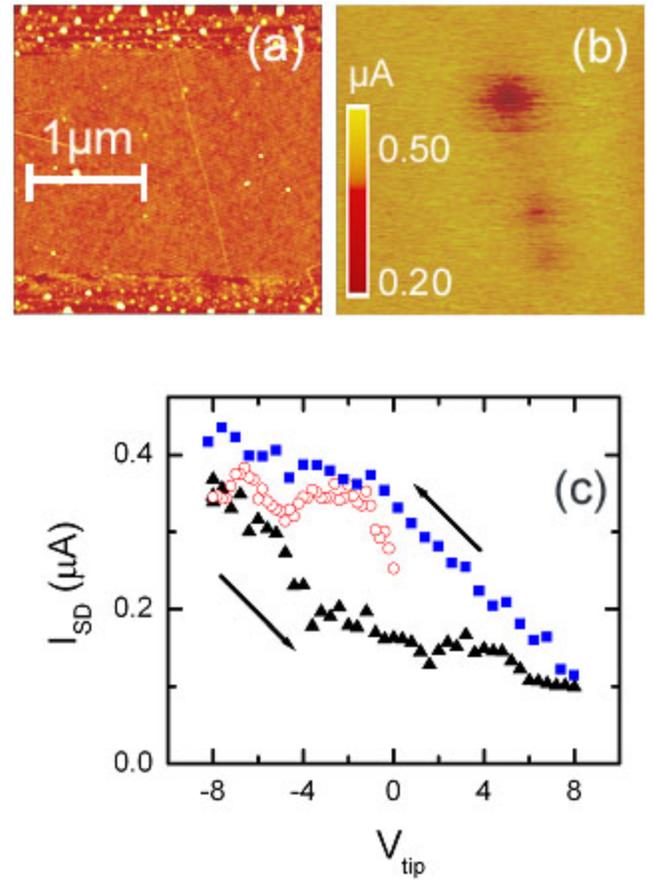

**Figure 2.** (a) AFM image of the swCN-FET presented in Figure 1. The diameter of the swCN is 1nm. (b) SGM image for the same device, with $V_{tip}$= +8 V, $V_{AC}$ = 100 mV, f = 10 kHz, $V_G$ = 0 V, and scan height h=15nm. The positive tip voltage depletes holes locally; the current $I_{SD}$ decreases when the tip is above defect sites. (c) Transport current vs. tip voltage with the SGM tip positioned above the top defect in Figure 2b. The bias voltage is 30 mV, and the backgate voltage is $V_G$ = 0 V. *A local* hysteresis (memory effect) is seen as the tip voltage is swept from positive to negative (blue squares, red circles) and back (black triangles). The memory effect is seen *only* with the tip above the SGM defect, which indicates that it can be induced at well-defined locations along the nanotube length.

In Figure 2b a positively biased tip ($V_{tip}$ = +8 V) depletes the carrier (hole)

concentration locally, and three spots (defects) are seen along the nanotube length. The diameter of the spots increases linearly with tip voltage, as previously found.[3,15] We observe a decrease in the transport current by a factor of 2.5 when the tip is above the top defect in comparison to the case when there is no tip gate.[16]

To further investigate the effect of the tip gating, we measure the dc transport current $I_{SD}$ as a function of $V_{tip}$ with the tip positioned at a fixed height (h = 15 nm) above the top defect in Figure 2b (Figure 2c). As the tip gate voltage is swept back and forth between +8V and –8V, a modulation in the transport current by a factor of 4 and a pronounced hysteresis is seen in $I_{SD}(V_{tip})$. A similar hysteresis ("memory effect") is commonly observed when the backgate voltage is swept, and was attributed to result from charge injection into stable trap states in the $SiO_2$ substrate due to the large electric field at the surface of the swCN-FET.[6,7] When the backgate voltage is swept, electric field-induced charge injection occurs all along the swCN-FET length. In contrast, the tip gate voltage induces a large electric field *locally* in a region of size roughly equal to the tip radius (30 nm). The size of the observed hysteresis loop due to the tip gate (8 – 9 V) is larger than that induced by the backgate (4 – 5 V). This is consistent with the fact that the small separation between tip and swCN leads to a maximum (local) electric field that is larger than occurs with the backgate at the same voltage.

There should be no memory effect if charge injection is localized to a region of the swCN-FET that shows no local gating response. In agreement with this prediction, we observe the memory effect of Figure 2c only if the tip is located over a defect region and never if it is located elsewhere over the swCN-FET. This is strong evidence that the active region for the memory effect can be as small as a few tens of nanometers. It should therefore be possible to miniaturize a nanotube-based memory element to less than 100 nm in size. Results of further investigations on the local memory effect will be discussed in detail elsewhere.[17]

Conventional Scanning Gate Microscopy measurements (Figure 2) provide information on variation of the low-frequency (DC – 10 kHz) resistance of the sample in response to the tip-gate. We have extended this technique to High-Frequency Scanning Gate Microscopy to investigate the effect of the defect regions on the swCN-FET impedance at frequencies up to 15 MHz. The data show that the circuit impedance may be modeled by the same parallel RC circuit model used to fit the data of Figure 1. The effect of the tip-gate is simply to increase the sample resistance, in agreement with the results of SGM measurements at low frequency.

In High-Frequency SGM, the voltage biased SGM tip is scanned over the swCN-FET at a fixed height (Figure 3c), and the circuit impedance modulus |Z| and phase θ, measured at a particular frequency f, are recorded as a function of tip position. Typical images are displayed in Figure 3a, and Figure 3b.

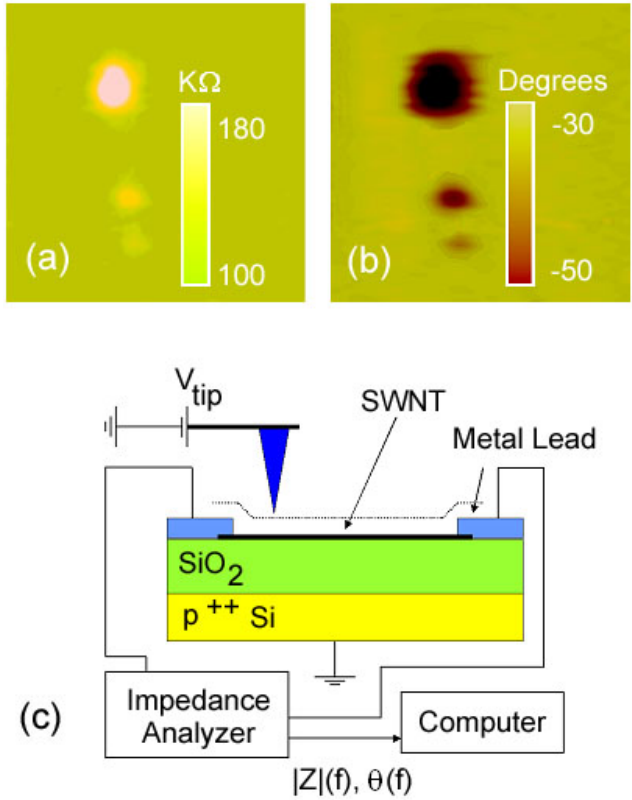

**Figure 3.** (a), (b) High Frequency Scanning Gate (HF-SGM) images, of |Z| (a) and θ (b) vs. tip position; data are taken for the same device presented in Figure 1 and Figure 2, for $V_{AC}$ =100mV, f = 300KHz, $V_{tip}$ = +8V, $V_G$ = 0V, scan height h = 15nm. The three defects seen in SGM (Figure 1b) are also present in these images. Changes in |Z| and θ when the tip is positioned above the defects are consistent with the predictions of a parallel RC circuit for the swCN-FET. (c) Schematic for HF-SGM.

Figure 3a and 3b show the variation of |Z| and θ, respectively, with tip position for $V_{tip}$= +8V, ac bias voltage $V_{AC}$ = 100 mV, and frequency f = 300 kHz. The same defect regions observed in SGM (Figure 2b) appear in HF-SGM images, implying that changes in the swCN-FET impedance are due to depletion of these regions by $V_{tip}$. The HF-SGM apparatus also allows a measurement of the swCN-FET impedance spectrum |Z(f)| with the SGM tip placed at h=15nm above the top defect, and $V_{tip}$=+8V (blue triangles in Figure 1). A fit to the data using a parallel R-C model yields a capacitance of 2.3pF and resistance of 310 kΩ for the circuit. The capacitance is the same as that found from conventional impedance spectroscopy measurements (red circles and green squares in Figure 1), while the resistance agrees with that found from the DC I($V_{tip}$) data in Figure 2c. We emphasize that these results show defects play a similar role at low and high frequencies, and the detailed response of |Z(f)| and θ(f) to the tip-gate voltage agrees with a simple parallel R-C model of the swCN-FET.

In conclusion, we have investigated the electronic properties of swCN-FETs at frequencies up to 15 MHz using Impedance Spectroscopy (IS) and High-Frequency Scanning Gate Microscopy (HF-SGM). We find that the IS data is well explained by a parallel RC circuit model where the resistance is set by the swCN-FET and the capacitance dominated by parasitics from the electrical connections. HF-SGM allows us to image precise locations along the swCN-FET length which respond strongly to the tip-gate and change the circuit resistance. The details of the response of |Z(f)| and θ(f) to the tip-gate voltage agree with the parallel RC model of the swCN-FET. This extension of SGM should prove useful in experiments on other systems including semiconductor nanowires, polymer nanofibers, and the like. Finally, we have demonstrated that the voltage-biased SGM tip-gate induces a strong memory effect only at specific locations along the length of the swCN-FET. This is strong evidence that swCN FET-based memory cells may be miniaturized to dimensions far below the micrometer scale of current devices.

**Acknowledgement.** We thank Yangxin Zhou for assistance with the shadow mask evaporation process and Sergei Kalinin for helpful discussions. This work was supported by the Laboratory for Research on the Structure of Matter (NSF DMR00-79909) and the National Science Foundation NIRT 0304531.

[*] E-mail: cstaii@physics.upenn.edu

**Acknowledgement.** We thank Yangxin Zhou for assistance with the shadow mask evaporation process and Sergei Kalinin for helpful discussions. This work was supported by the Laboratory for Research on the Structure of Matter (NSF DMR00-79909) and the National Science Foundation NIRT 0304531.

[*] E-mail: cstaii@physics.upenn.edu

(15) An analysis similar to that used in ref 3 gives 19 meV for the Fermi energy at zero gate voltage for the top (strongest) defects seen in Figure 2b. This value is within 10% of that found for the strongest defects in ref 3.

(16) Since the swCN-FET ON/OFF ratio induced by a positive backgate voltage exceeds 1000, it is at first surprising that the current decreases by only a factor of 2.5 with +8 V applied to the SGM tip. There are at least two explanations for this. First, the tip gate depletes only a small fraction of the swCN length (~ 30 nm) while the backgate couples capacitively to the entire sample. Therefore the backgate voltage simultaneously affects *all* defects along the nanotube length *and* the Schottky barriers at the contacts to the swCN-FET, which together are responsible for the large measured ON/OFF ratio. Second, according to the calculations of ref 18, the local potential at the nanotube is only a 0.1 – 1 % of the SGM tip voltage for scan heights 3 to 50 nm. Therefore, with $V_{tip}$ = +8 V the local potential at the defect site is less than 80 mV. The defect region will be fully depleted (so the current suppressed to zero) only if the local potential exceeds V*, the defect depletion surface potential (ref 3). From SGM measurements and an analysis as outlined in ref 3, we find V* = 230 mV for the top defect in Figure 2b, a value much larger than the local potential at the defect site due to the tip gate.